\documentclass[1p]{dbarticle1}

\usepackage{graphicx}
\usepackage{natbib}
\usepackage[normalem]{ulem}

\begin{document}

\title{The Discrepancy in Modelling of Galactic Cosmic Ray Primaries and Secondaries}

\author{D. Bisschoff}\corref{cor1}
\author{I. B\"usching}
\author{M. S. Potgieter}

\cortext[cor1]{Corresponding author}

\address{Centre for Space Research, North-West University, Potchefstroom 2520, South Africa}


\begin{abstract}
Given that the bulk of the galactic cosmic rays (CRs) originates from transient point-like sources, such as supernova remnants, the flux of the CR primary component measured at Earth depends on the local source history. Whereas the secondary component shows little to no variation due to nearby sources, and instead, should depend on the global distribution of CRs. Steady-state, rotational symmetric models (2D) of CR propagation assume smeared-out CR sources in the Galaxy and cannot properly take into account the influence of nearby point sources. In this paper, we infer evidence of nearby sources by illustrating that a 2D propagation model will not describe CR primaries and secondaries equally well. We adapted the 2D version of the GALPROP code to a compute-cluster environment and perform parameter studies comparing CR spectra with primary and secondary CR data separately. Doing a parameter study, one may expect different best fit values looking at the primary and secondary CR components separately, as it is 
unlikely that the source history mimicked by the 2D models coincides with the real local source history. We find that the primaries and secondaries are fitted with differing best fit parameters, with the results indicating that the 2D model being more suited to modelling secondary CRs. The insufficient modelling of primary CRs can be contributed to the possible presence of local CR sources. 
\end{abstract}


\maketitle

\newpage
\section{Introduction}
Calculations have shown that the flux of the cosmic ray (CR) primary component measured at Earth depends on the local source history, given that the bulk of the galactic CRs originates in transient, point-like sources \citep{Busching2005} like supernova and their remnants. The secondary CR component is not directly affected by local point-like sources and instead should depend on the global distribution of CRs. This suggests that the widely used steady-state, rotational symmetric models (2D models) of CR propagation \citep{Moskalenko2003,Ptuskin2006,StrongMoskalenko1998,Strong2000} might not adequately describe the CR primary component originating from transient, point sources, but are more suited to model the secondary component. These 2D models assume smeared-out sources, which do not necessarily result in the same local CR flux as the real local sources would, leading to an insufficient description of the CR primary component. When working with 2D models, concentrating on secondary, tertiary and higher CR 
nuclei seperately, may thus yield a better description of the galactic CR propagation.

In this paper we test the assumption that steady-state 2D CR propagation models (e.g. the one incorporated in the GALPROP code) are beter suited for the CR secondary component than the CR primary component. In order to do so we perform a parameter study, obtaining best fit parameters for the CR primary and secondary components separately. We interpret our results as an indication for nearby CR sources in the chemical composition of the local CR flux.

\section{Method, Assumptions and Calculations}
For CR propagation, the Galaxy can be described as a cylinder with a radius of $\approx$ 20\,kpc and a height of up to $\approx$ 4\,kpc, including the galactic halo, in which CRs have a finite chance to return to the galactic disk. Assuming symmetry in azimuth leads to 2D models that only depend on Galactocentric radius and height. Neglecting the time dependence leads to a steady-state model. Time-dependent calculations taking into account all three spatial dimensions are still numerically too involved for large parameter studies, so the 2D version of the GALPROP\footnote{http://galprop.stanford.edu/web\_galprop/galprop\_home.html} code \citep{Moskalenko2003,StrongMoskalenko1998,Strong2000} is used for this very extensive parameter study.

For results presented here we used the plain diffusion model \citep{Ptuskin2006} as implemented in GALPROP. This model was chosen because it doesn't take reacceleration into account which simplifies the model and reduces the number of free parameters to consider. Although reacceleration is an important process, it is not considered dominating for energies higher than 1\,GeV/nuc \citep{StrongReview2007}. As this study was limited to energies above 4\,GeV/nuc, we believe the impact of reacceleration would have on our final conclusions is minimal. Limiting the parameter studies to higher energies ($>$ 10\,GeV/nuc) to have even less contribution from reacceleration would not have been feasible as it would have lowered the amount of available data points significantly.

The plain diffusion model was used in cylindrical coordinates with two spatial dimensions, the galactocentric radius $r$ and the halo height above the galactic plane $z$, with symmetry in the angular dimension. The halo height was fixed to 4\,kpc, as generally used in these studies. Varying the size of the halo can be counteracted by directly varying the diffusion coefficient, thus the halo was kept constant. Initial tests have shown that the galactic wind gives only minor changes in the fluxes for rigidity above 4\,GV. The velocity and gradient in the galactic wind can thus be set to zero, simplifying the model. The energy range over which the model was run was kept to the default range, but this study considered only data values above 4\,GeV/nuc. The effect of solar modulation is lower at these high energies thus any errors made in describing the modulation have less impact on the final results. All further parameters in the model, such as source abundance values and interstellar properties, are adopted 
from \citet{Ptuskin2006}. These include the cross sections and gas densities. Gas densities are defined as cylindrically symmetrical distributions for H2, HI and HII. As we are interested in studying primaries and secondaries separately, the choice of both cross-sections and gas density will affect the results obtained. Discussion of the fundamental dependence of our results on these parameters are beyond the scope of this study and will to be addressed in future studies.

\begin{table}[!h]
\footnotesize
\centering
\caption{Parameter space considered.}
\label{param}
\begin{tabular}{llll}
\hline
 Parameter & Min &Max &Unit\\\hline
$k_0$&0.50&  5.0 &$ 10^{28}\,{\rm cm}^2{\rm s}^{-1}$\\
$\delta$ &  0.1 & 1.0& \\ 
$\alpha$ & 1.50 & 3.50&\\ 
\end{tabular}
\end{table}

We scanned the parameter space given in Table \ref{param}. Here $k_0$ determines the magnitude of the diffusion coefficient ($D_{xx} = \beta k_0 (\rho / \rho_0)^{\delta}$) at reference rigidity of 4\,GV and $\delta$ the power index of the energy dependance of the diffusion coefficient. Additionally $\alpha$ is the spectral index of the sources. While $\alpha$ directly affects the primary CRs, it indirectly affects secondary CRs as the secondaries are dependant on the primaries. This range was chosen as to vary the free parameters over a wide range of possible values, but also to include the values obtained by \citet{Ptuskin2006}. No breaks in $\alpha$ or $\delta$ were implemented at low energies.

\begin{table}[!ht]
\footnotesize
\centering
\caption{List of experimental data sets used and the corresponding estimated force field parameters.}
\label{ffp}
\begin{tabular}[t]{ l  l  l  l }
\hline
Experiment &  Reference   & Parameter (MV) & CR species\\ \hline
AMS01    & \citet{Ag02} & 680  &  $^4$He\\
ATIC2    & \citet{Pa06} & 885  &  $^4$He, He, C, O, CNO, Ne, Mg, Si, Fe\\ 
BESS     & \citet{Sa00} & 750  &  $^4$He\\
CAPRICE98& \citet{Bo03} & 950  &  $^4$He\\ 
CRN      & \citet{Mu91} & 700  &  C, O, Ne, Mg, Si, Fe\\ 
         & \citet{Sw93} &      &  C+N+O, Ne+Mg+Si\\
HEA03    & \citet{En90} & 885  &  Be, B, C, N, O, F, Ne, Na, Mg,\\
         &              &      &  Al, Si, P, S, Cl, Ar, K, Ca,\\
         &              &      &  Sc, Ti, V, Cr, Mn, Fe, Co\\
IMAX     & \citet{Me00} & 750  &  $^4$He\\ 
JACEE    & \citet{As98} & 900  &  $^4$He, C+N+O, Ne+Mg+Si, Fe\\ 
MUBEE    & \citet{Za94} & 700  &  $^4$He\\ 
RUNJOB   & \citet{De05} & 885  &  $^4$He, C+N+O, Ne+Mg+Si, Fe\\ 
SANKIRU  & \citet{Ka97} & 700  &  Fe\\ 
SOKOL    & \citet{Iv93} & 700  &  $^4$He, C+N+O, Ne+Mg+Si, Fe\\ 
\hline
\end{tabular}
\end{table}

\newpage
A total of 30720 models were calculated and the calculations were performed on the institutional cluster of the North-West University in Potchefstroom using a MPI\footnote{Message Passing Interface: http://mpi-forum.org/} code to run the models in parallel. 

The full nuclear reaction network was solved over all isotopes implemented in GALPROP. Thus both primary and secondary CR species were run at the same time. The LIS calculated with GALPROP were then compared to data by method of a $\chi^2$ test in order to find the best fit parameter set, computed from the CR database\footnote{http://www.mpe.mpg.de/$\sim$aws/propagate.html} \citep{Database}. Protons have been used in this study to normalise the computed spectra to data and account for modulation. For each model, we calculated the $\chi^2$ value for each entry in the database when compared to the corresponding calculated LIS value after the temporal variation of the modulation during a solar cycle was taken into account. At energies $<$ 10\,GeV/nuc the effect of solar modulation has to be considered. This was done by using the force field approximation \citep{Gleeson1968a,Gleeson1968b} with a set of modulation parameters (as listed in Table \ref{ffp}), obtained by comparing a proton LIS to proton data from 
different epochs in the solar cycle. While being a simple model it describes the solar modulation well enough for the purpose of this paper at the energy range considered.

Testing the assumption that 2D models are well suited to describe the CR secondary component, but are less effective in describing primary CR, we divide the existing CR data into three components according to the fraction of secondary nuclei they contain. The fraction of secondaries and primaries in the isotopes that make up each species are added up separately. The addition is weighted according to the known abundances of the isotopes in a species when detected at Earth, integrated over all energies \citep{Shapiro1991}. Using this method to differentiate between CR species, all the CR data and local interstellar spectra (LIS) can then be divided into one of three component groups: Primary CRs, Mixed CRs and Secondary CRs:
\begin{itemize}
\item Primary component: secondary fraction $<$30\%
\item Mixed component: secondary fraction $>$30\%,$<$70\% 
\item Secondary component: secondary fraction $>$70\%  
\end{itemize}
The resulting $\chi^2$ values were then added up for each of the three CR components (as distributed in Table \ref{crlist}) separately.

\begin{table}[!htb]
\footnotesize
\centering 
\caption{Species composition of the three CR component groups}
\label{crlist}
\begin{tabular}[t]{lll}
\hline
Primary component: & Mixed component: & Secondary component: \\ \hline
H        & N     & Be\\
He       & C+N+O & B\\ 
C        & Ne    & F\\  
O        & Na    & P\\  
Mg       & Al    & Sc\\ 
Si       & S     & Ti\\
Ne+Mg+Si & Cl    & V\\ 
Ca       & K     & Mn\\ 
Fe       & Cr    & \\
         & Co    & \\
\end{tabular}
\end{table}

\newpage
\section{Results}

\begin{figure}[!ht]
   \centerline{\includegraphics[width=2.2in, trim=40 20 10 20,clip=true]{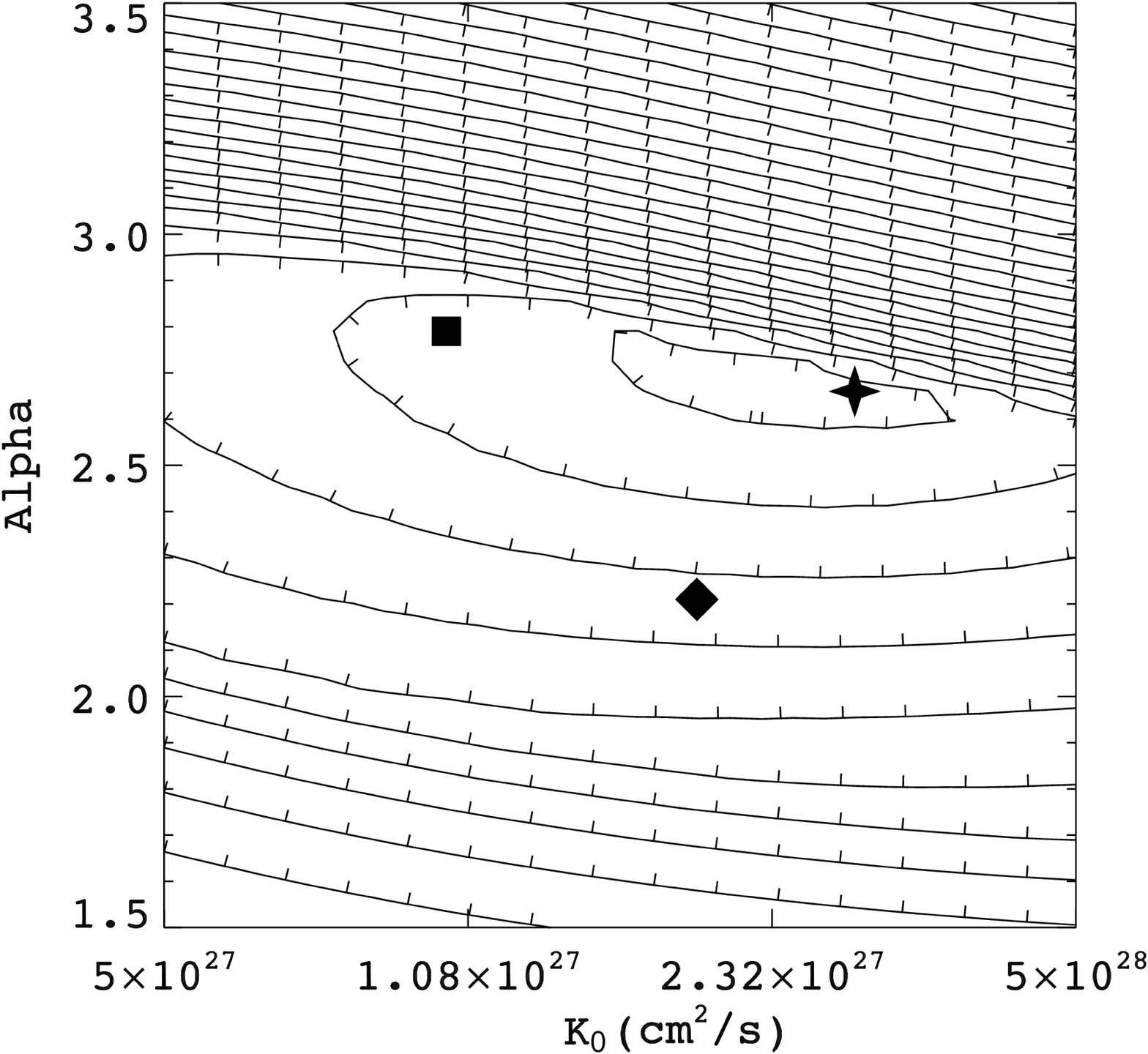}
              \includegraphics[width=2.2in, trim=40 20 10 20,clip=true]{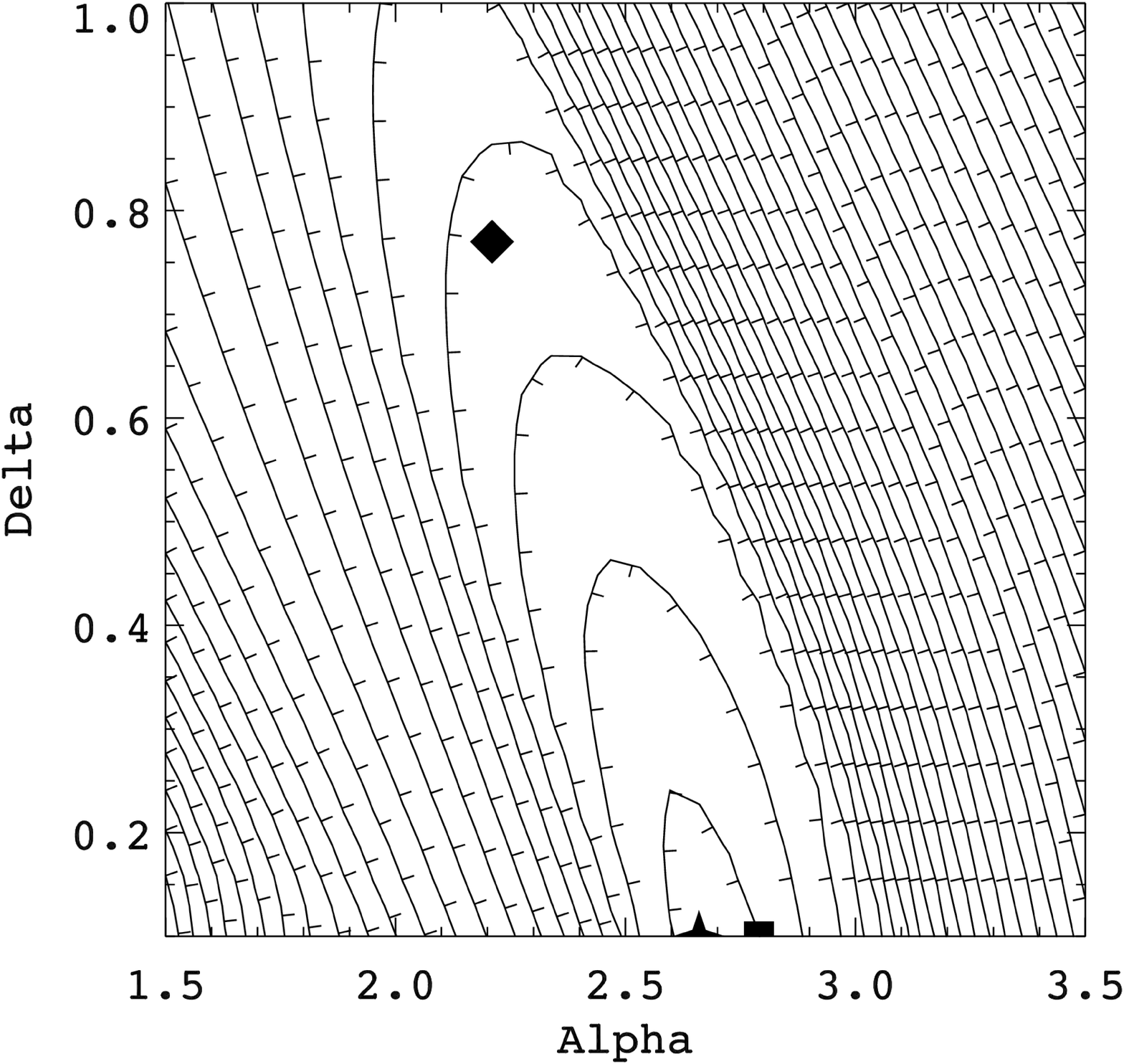}
              \includegraphics[width=2.2in, trim=40 20 10 20,clip=true]{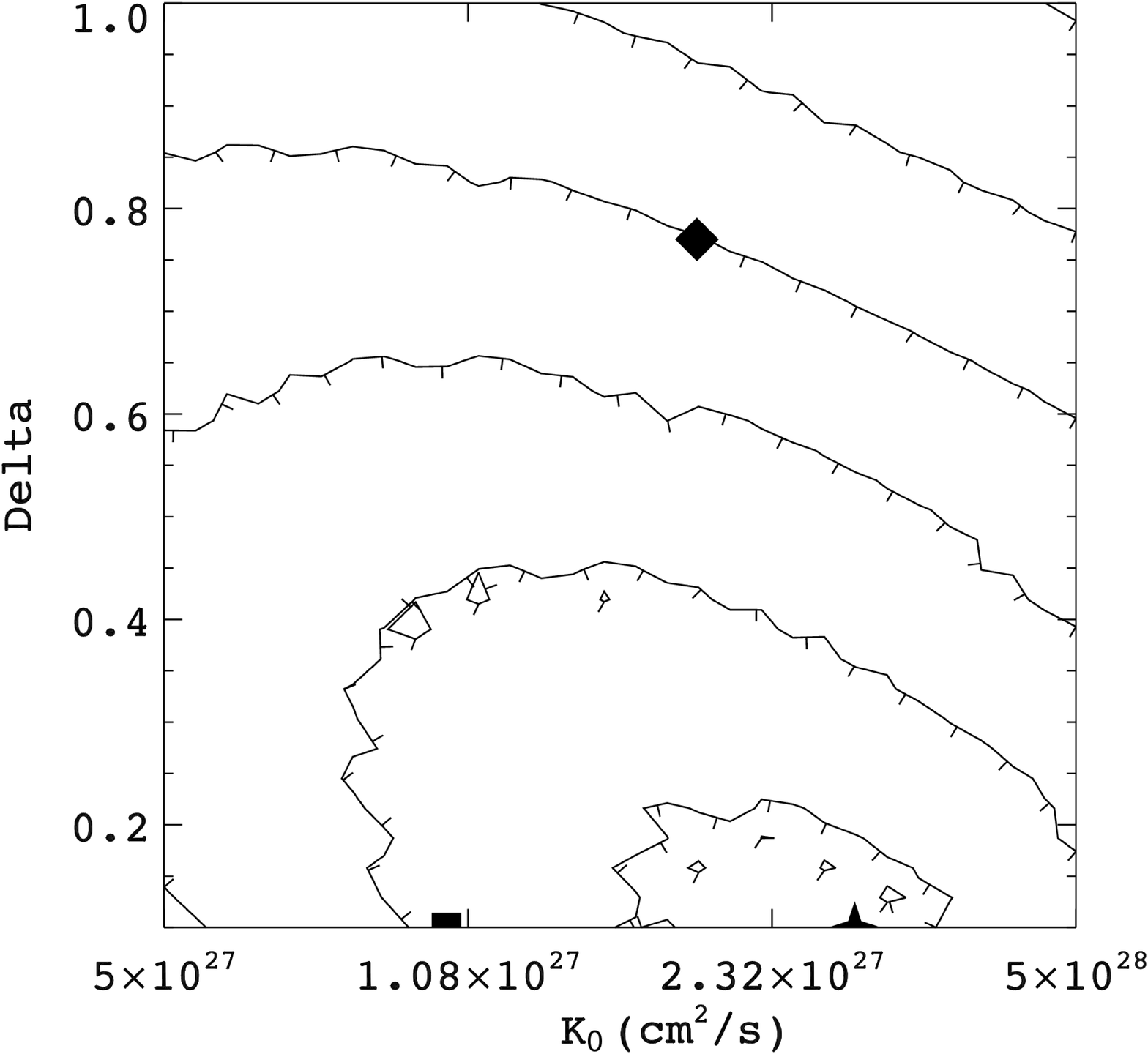}
             }
   \caption{$\chi^2$ distribution for the Primary CR component, when fitting the models to data above 4\,GeV/nuc, in the $k_0-\alpha$ (left)  $\alpha-\delta$ (middle) and $k_0-\delta$ (right) plane. Minimum value in each plane is marked by a 4-point star. The minimums for the other two components are marked for comparison, a diamond for the Secondary component and a square for the Mixed component.}
   \label{pri1}
   \centerline{\includegraphics[width=2.2in, trim=40 20 10 20,clip=true]{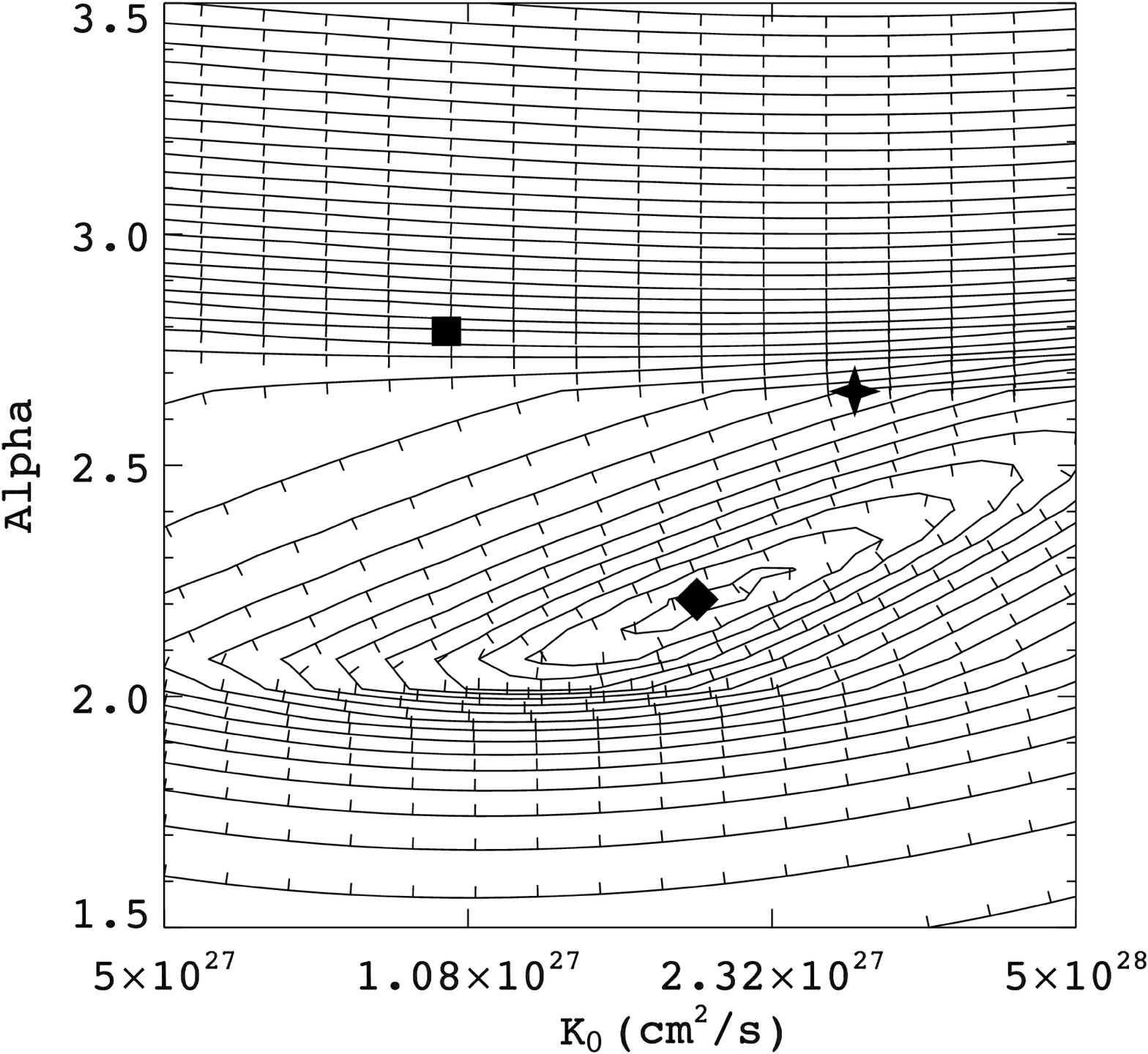}
              \hfil
              \includegraphics[width=2.2in, trim=40 20 10 20,clip=true]{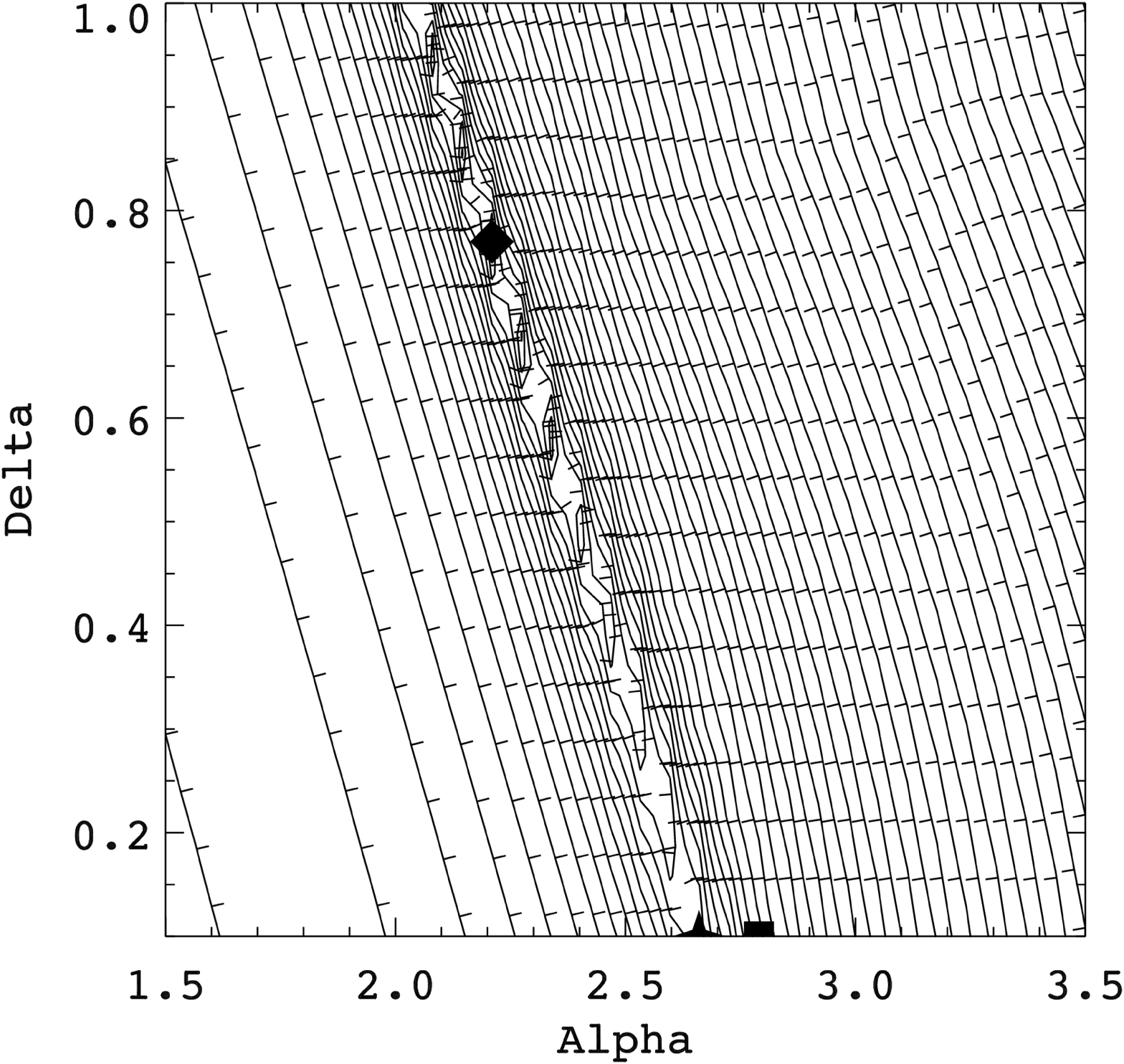}
              \hfil
              \includegraphics[width=2.2in, trim=40 20 10 20,clip=true]{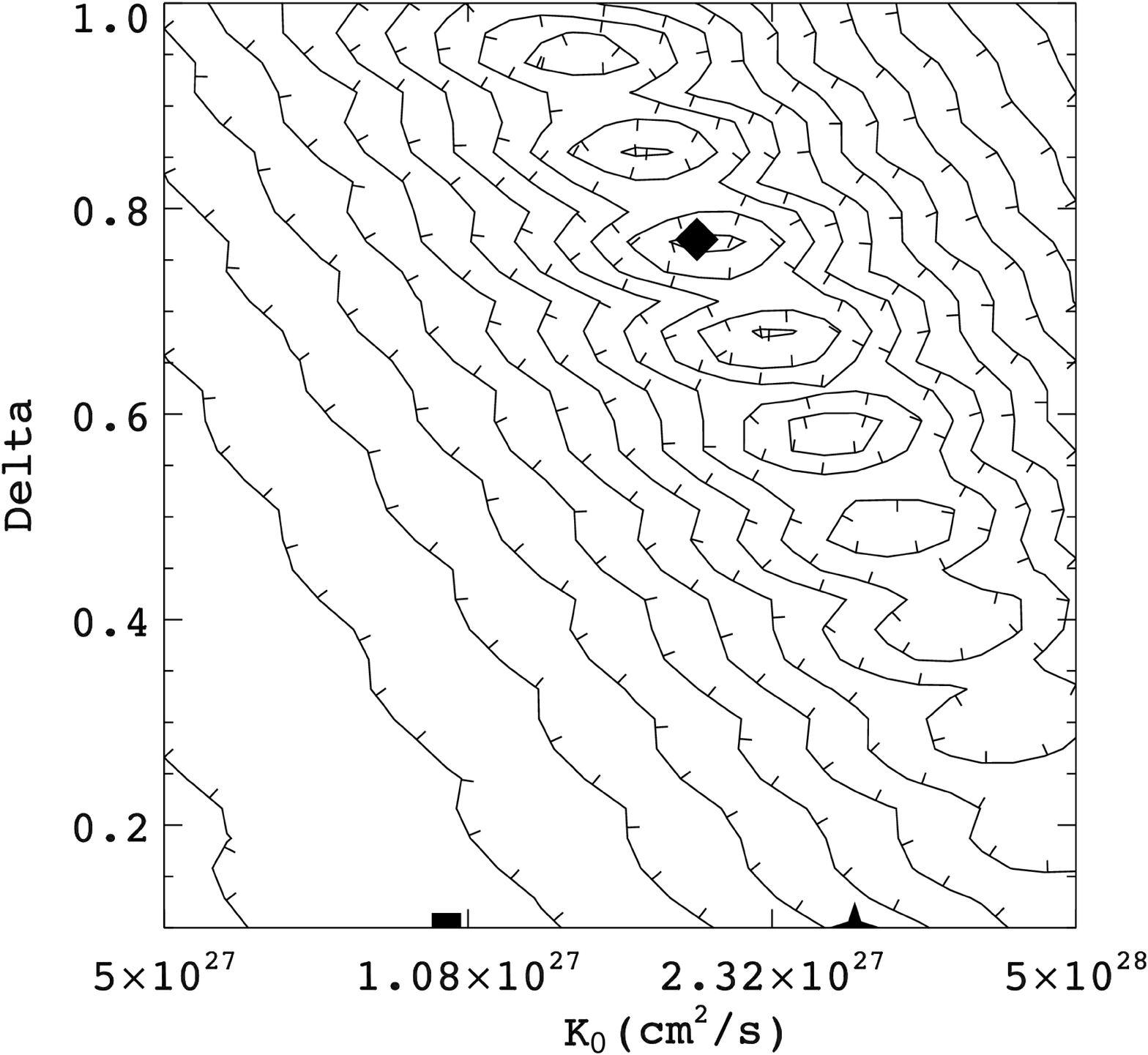}
             }
   \caption{Same as Fig. \ref{pri1}, but the $\chi^2$ distribution is for the Secondary CR component.}
   \label{sec1}
   \centerline{\includegraphics[width=2.2in, trim=40 20 10 20,clip=true]{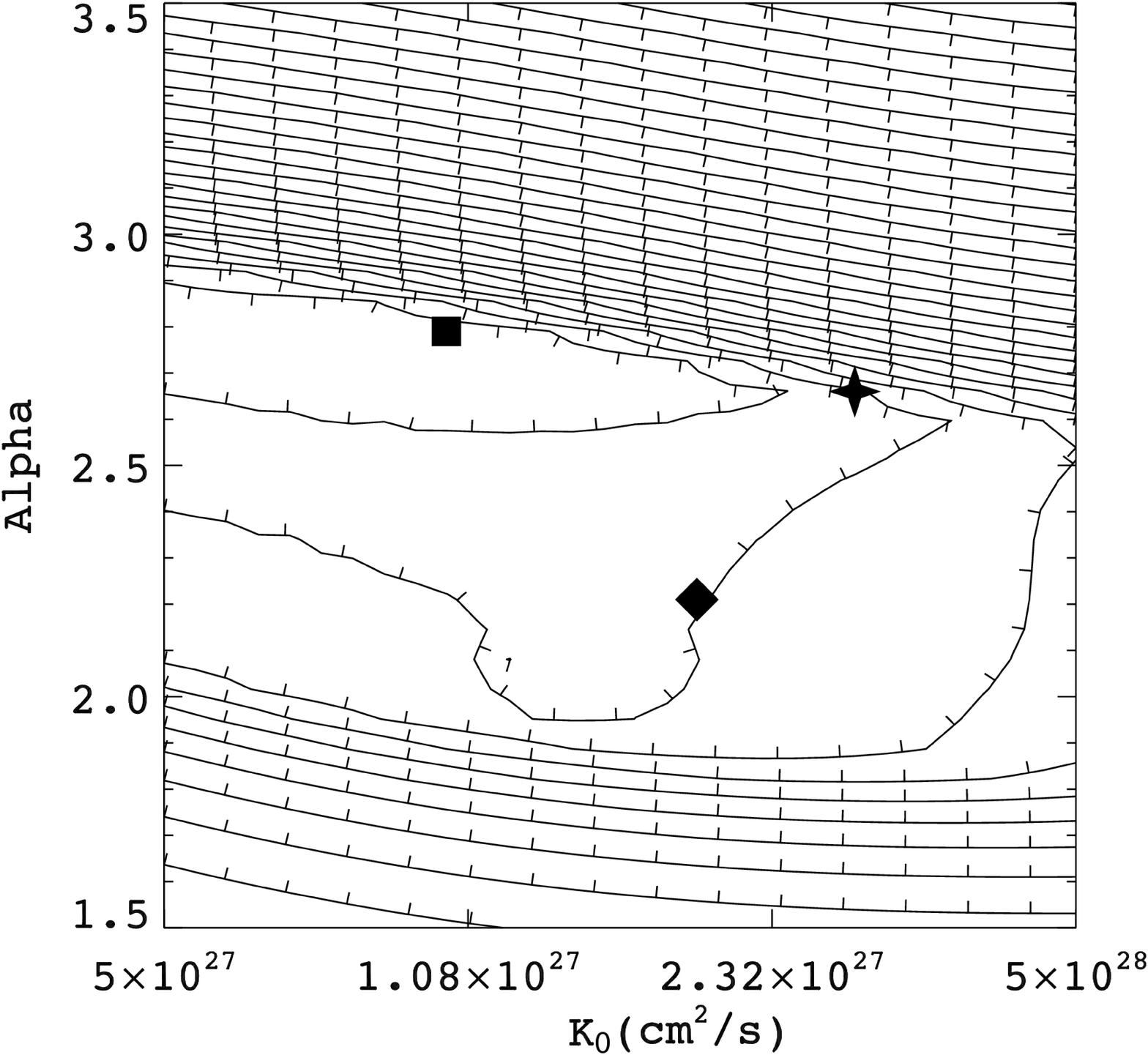}
              \hfil
              \includegraphics[width=2.2in, trim=40 20 10 20,clip=true]{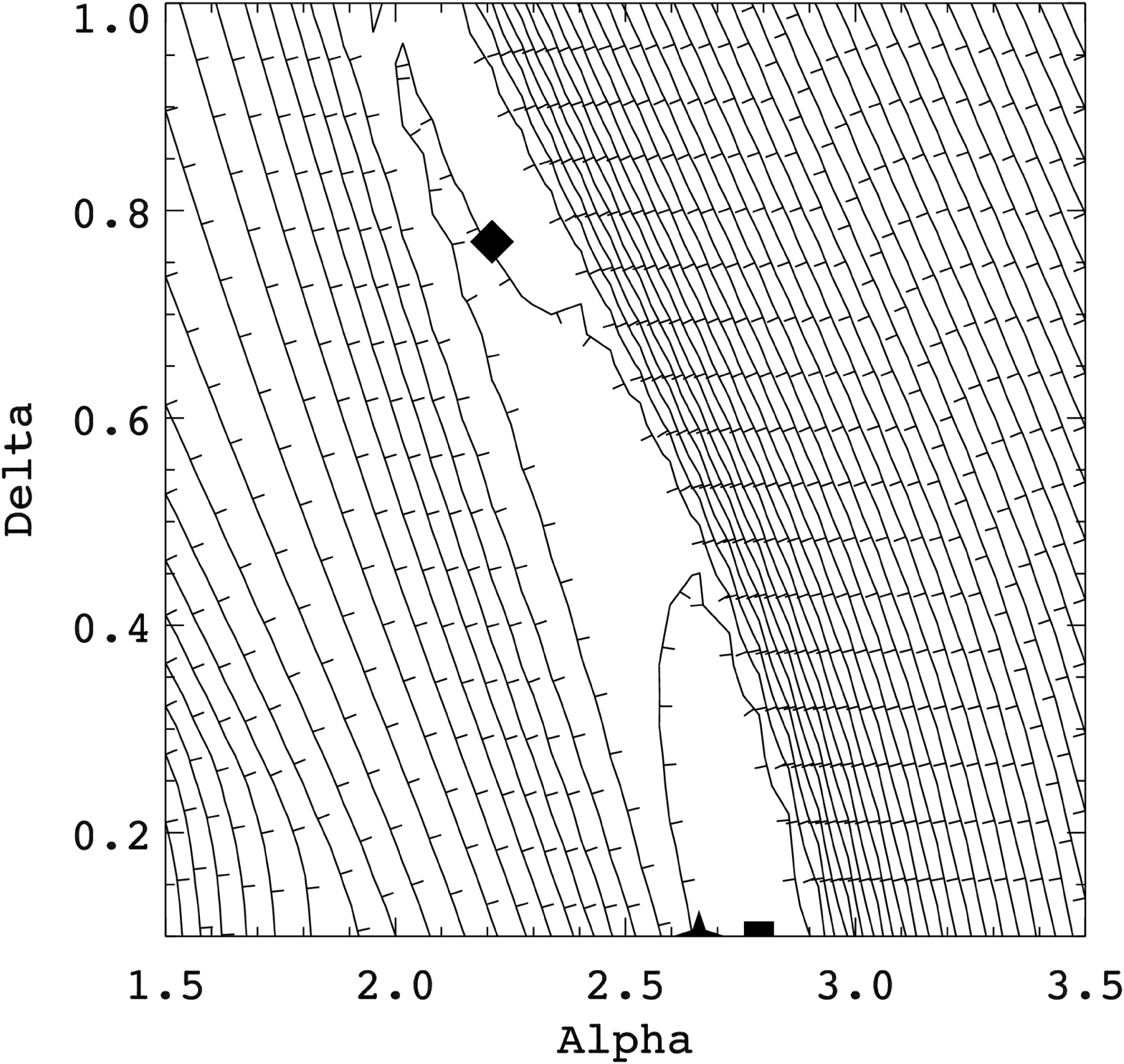}
              \hfil
              \includegraphics[width=2.2in, trim=40 20 10 20,clip=true]{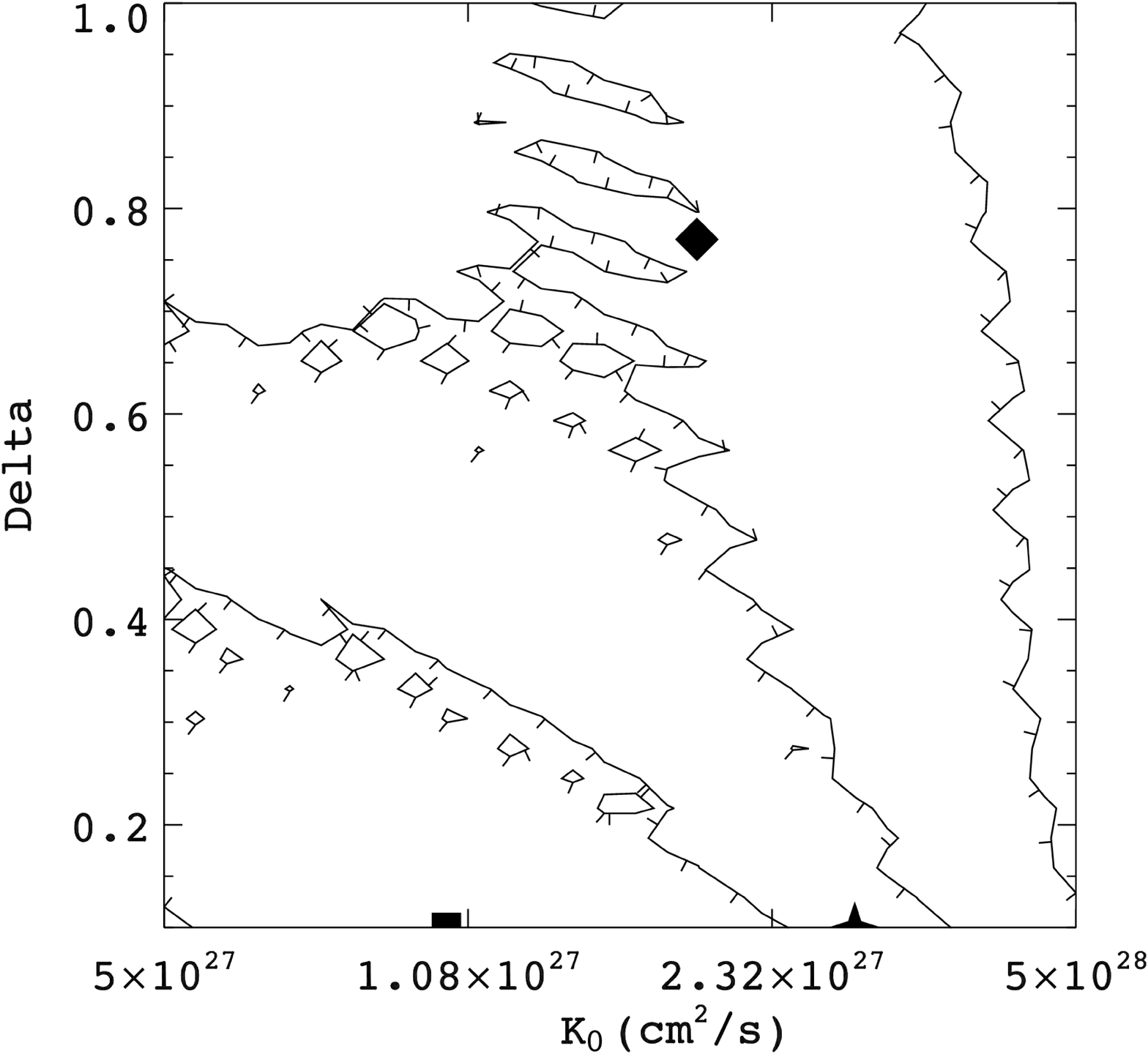}
             }
   \caption{Same as Fig. \ref{pri1}, but the $\chi^2$ distribution is for the Mixed CR component.}
   \label{mix1}
 \end{figure}

The results of our calculations are presented in Fig. \ref{pri1} to \ref{mix1}, where we show contour plots of best $\chi^2$ values over the parameter range considered for the CR primary, secondary and mixed component, respectively. Our best fit parameters for the three components are given in Table \ref{result} with the step size for $\alpha$: 0.0625 (linear), for $\delta$: 0.028125 (linear) and logarithmic for $k_0$ with factor of $10^\frac{1}{29}$. These best fit parameters are marked on the contour plots for easier comparison of the relative locations.

\begin{table}[!h]
\footnotesize
\centering
\caption{Best fit values for the secondary, primary and mixed components.}
\label{result}
\begin{tabular}{lllll}
\hline
 Parameter & Secondary & Primary & Mixed & Unit\\\hline
$k_0$& 1.92831 & 2.86808  & 1.02168  & $ 10^{28}\,{\rm cm}^2{\rm s}^{-1}$\\
$\delta$ & 0.767742  & 0.10000 &0.10000 & \\ 
$\alpha$ & 2.20968 &2.66129 & 2.79032  &\\ 
\end{tabular}
\end{table}

Considering the results shown in Table \ref{result} and Fig. \ref{pri1} and \ref{sec1}, the different locations of the minimum $\chi^2$ for primary and secondary CR component in the $k_0$-$\alpha$, $\alpha$-$\delta$, and $k_0$-$\delta$ planes is apparent. The $\chi^2$ contours are also quite different, thus the calculated spectra of the three components show different sensitivities to the model parameters.

Not surprisingly, the plots for the mixed component resemble somewhat a superposition of the corresponding secondary and primary plots. The high $\chi^2$ values for models with $\alpha$ not in the range $2.0 < \alpha < 3.0$ indicate that values outside this range can be disregarded.

In our calculations, the primary and secondary components of the galactic CR seem to favour different regions in the scanned parameter space. Additionally the best fit parameters obtained for primaries result in poor fits and very large $\chi^2$ values when used to fit the secondaries, similarly for using the primaries' best fit parameters to fit the secondaries. As mentioned in the introduction, a possible explanation of our findings is that 2D models are indeed incapable of correctly describing the contributions of local CR point sources.

\begin{figure}[!ht]
   \centerline{\includegraphics[width=2.8in, trim=1 15 1 27, clip=true]{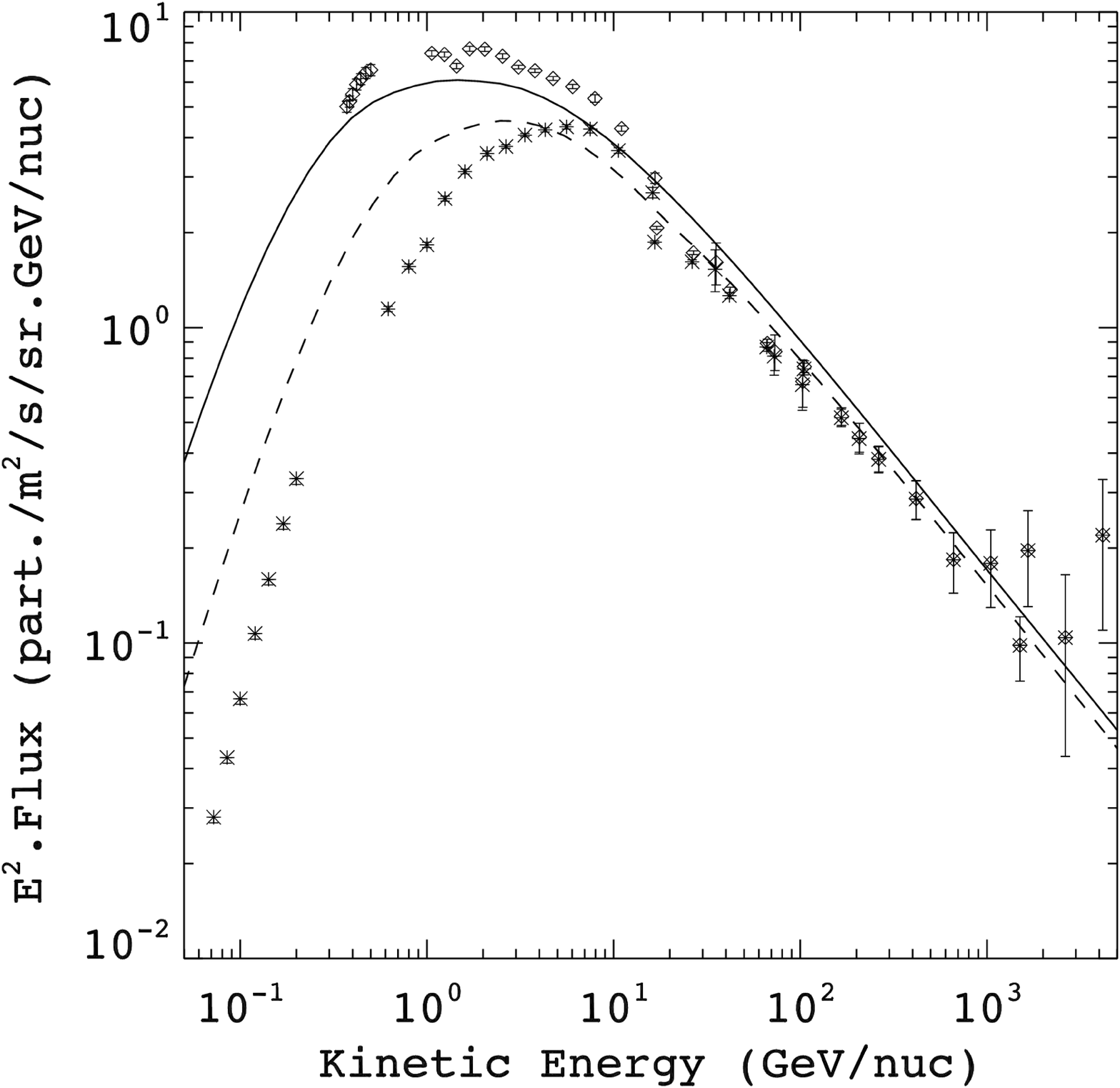}
              \hfil
              \includegraphics[width=2.83in, trim=1 15 1 27, clip=true]{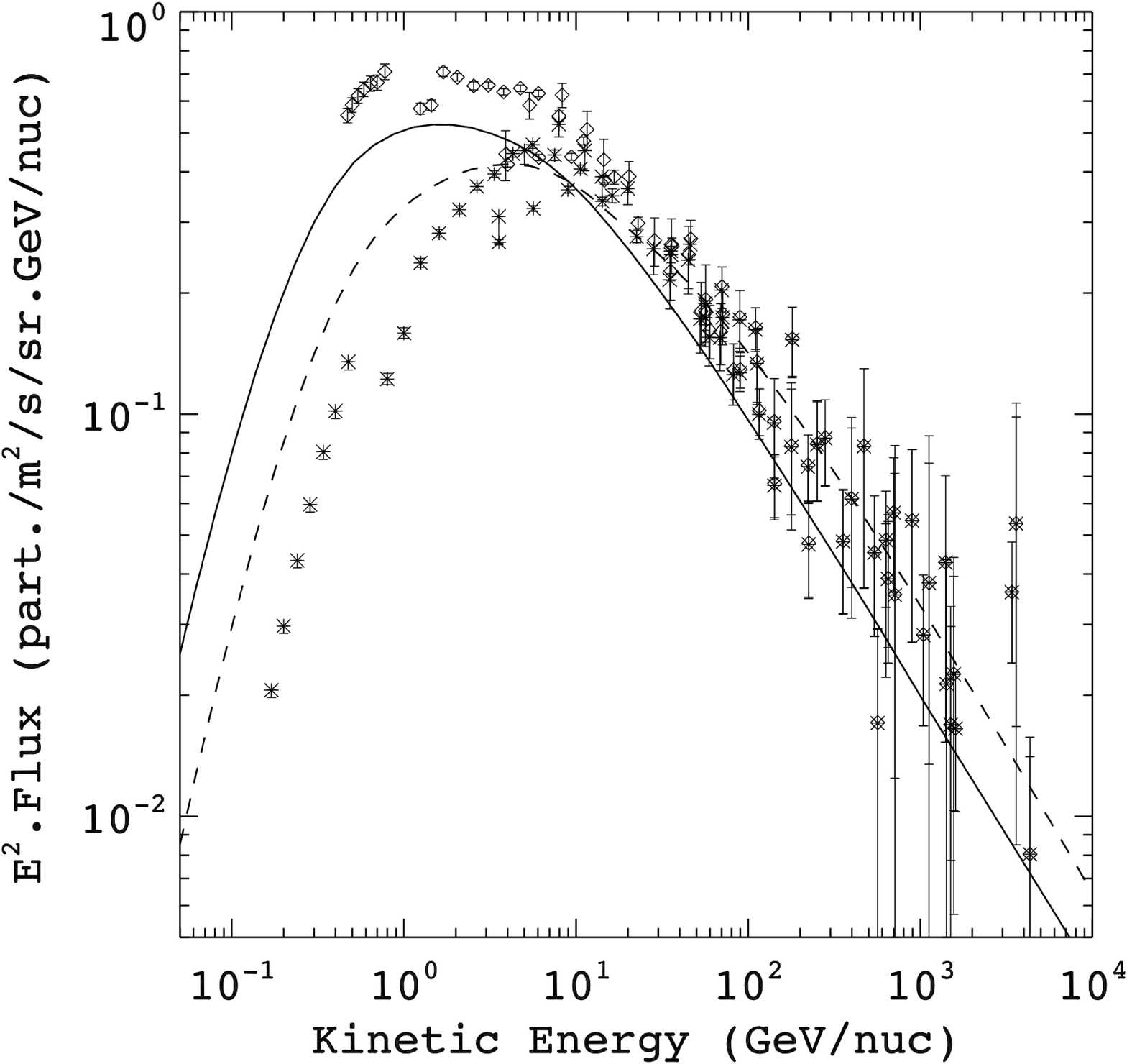}
             }
   \caption{LIS for Primary CR component species Carbon and Iron. The solid line is the LIS for this study and the dashed line is the LIS from \citet{Ptuskin2006}. Experimental data is marked with stars and the data with the effect of solar modulation removed is marked with diamonds. Only data points above 4\,GeV/nuc were used in the fitting.}
   \label{pri2}
   \centerline{\includegraphics[width=2.8in, trim=1 15 1 27, clip=true]{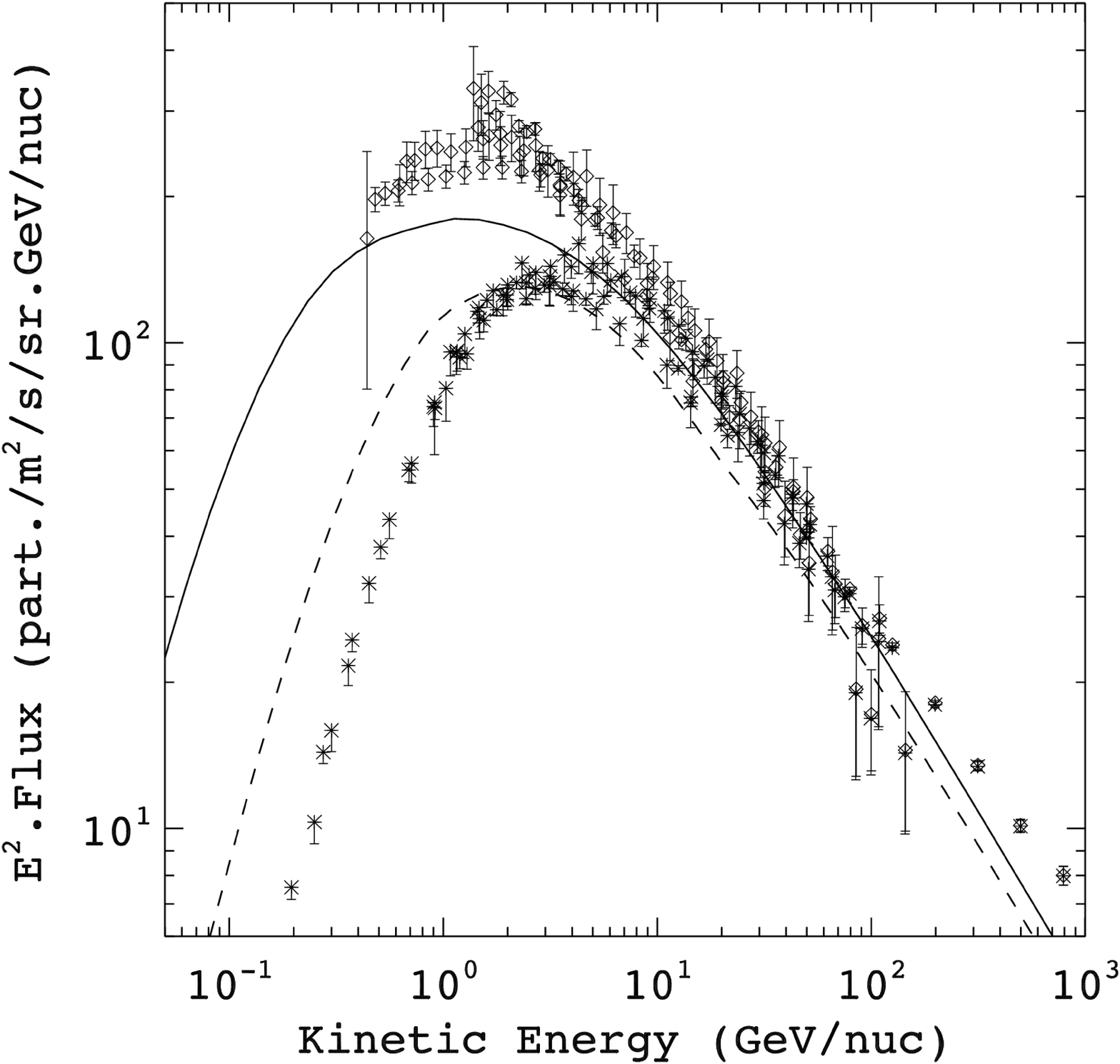}
              \hfil
              \includegraphics[width=2.8in, trim=1 15 1 27, clip=true]{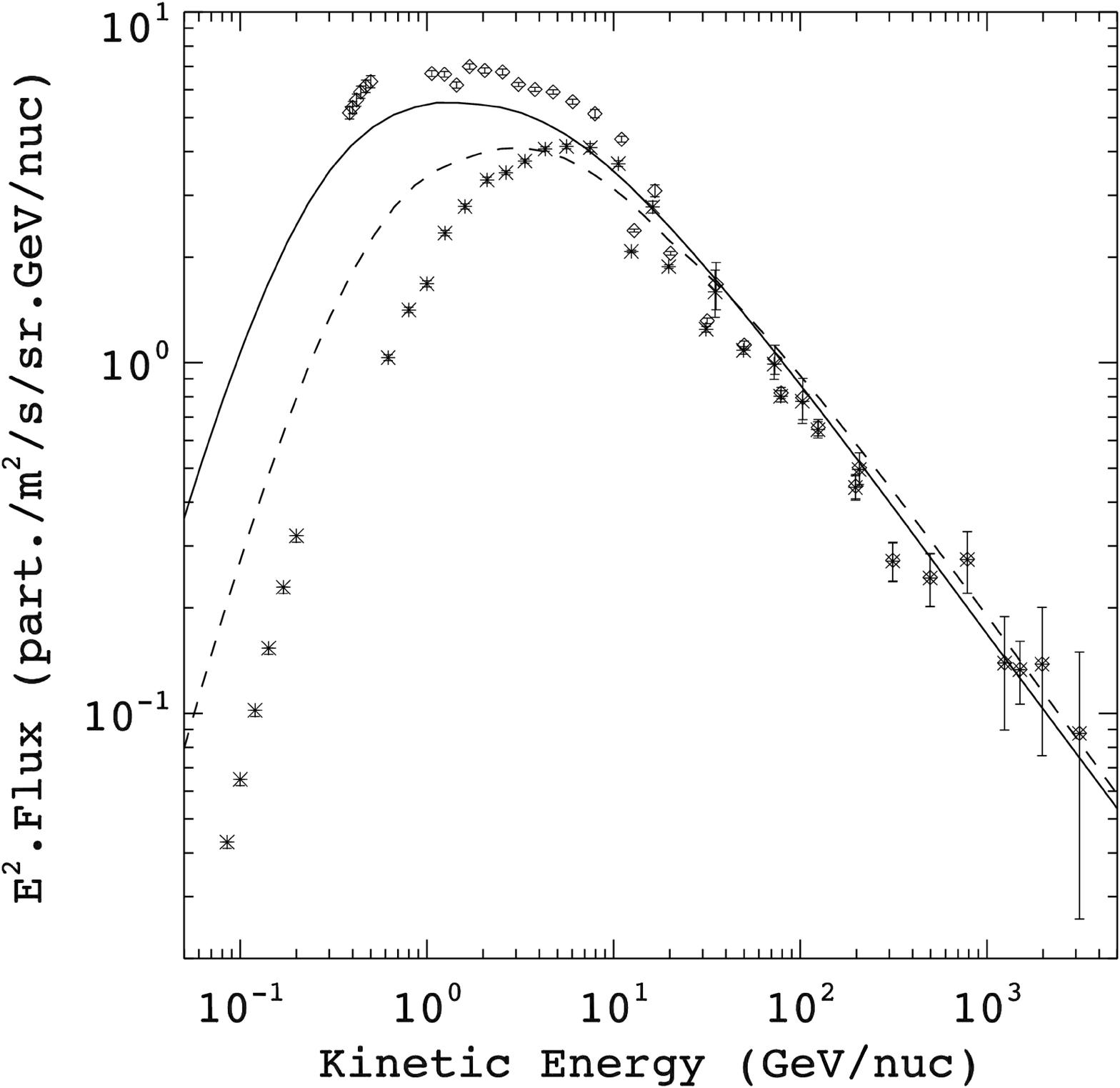}
             }
   \caption{Same as Fig. \ref{pri2} but the LIS for Primary CR component species Helium4 and Oxygen is shown.}
   \label{pri3}
\end{figure}

\begin{figure}[!ht]
   \centerline{\includegraphics[width=2.8in, trim=1 15 1 27, clip=true]{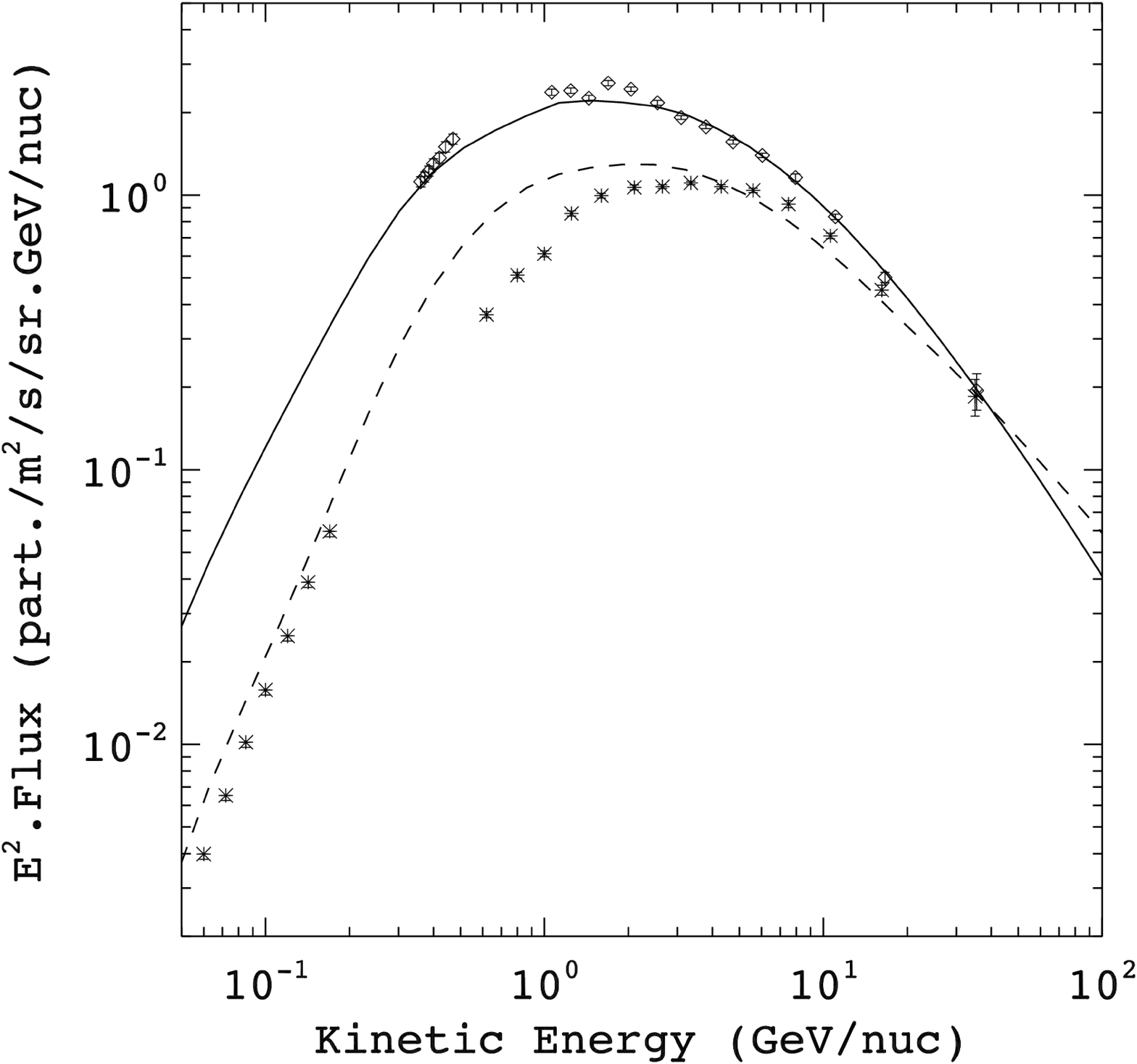}
              \hfil
              \includegraphics[width=2.8in, trim=1 15 1 27, clip=true]{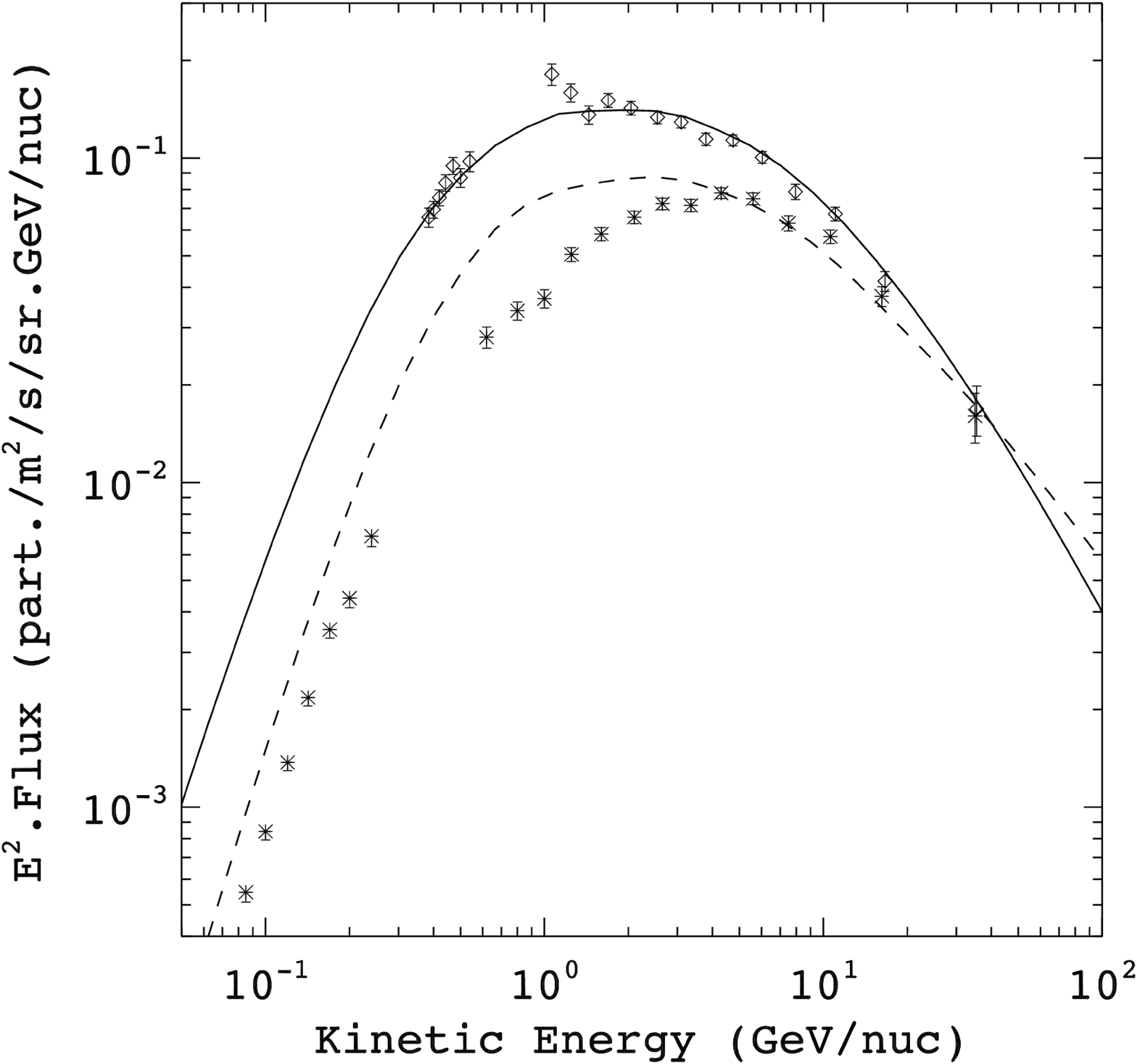}
             }
   \caption{Same as Fig. \ref{pri2} but the LIS for Secondary CR component species Boron and Fluorine is shown.}
   \label{sec2}
   \centerline{\includegraphics[width=2.8in, trim=1 15 1 27, clip=true]{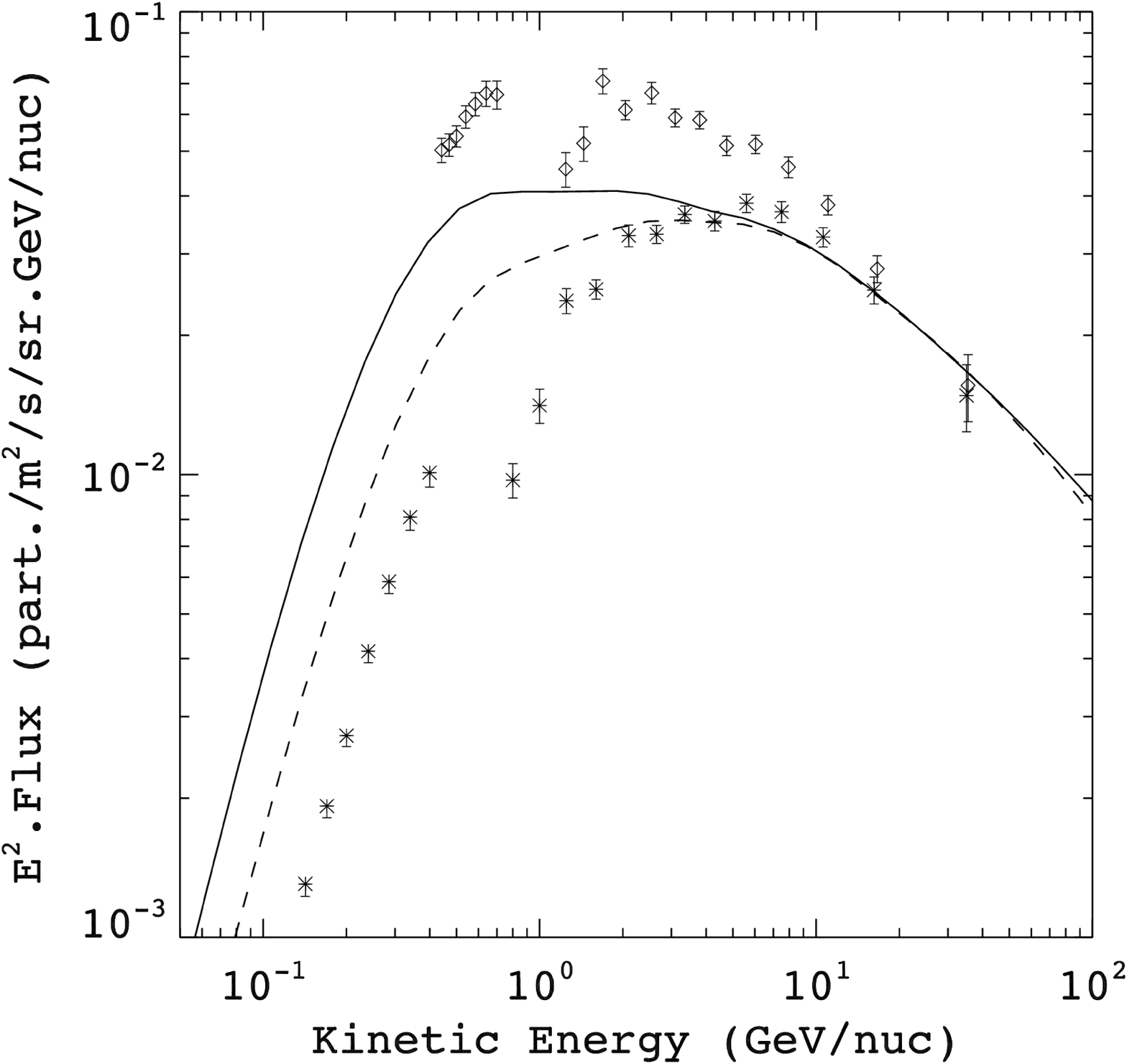}
              \hfil
              \includegraphics[width=2.8in, trim=1 15 1 27, clip=true]{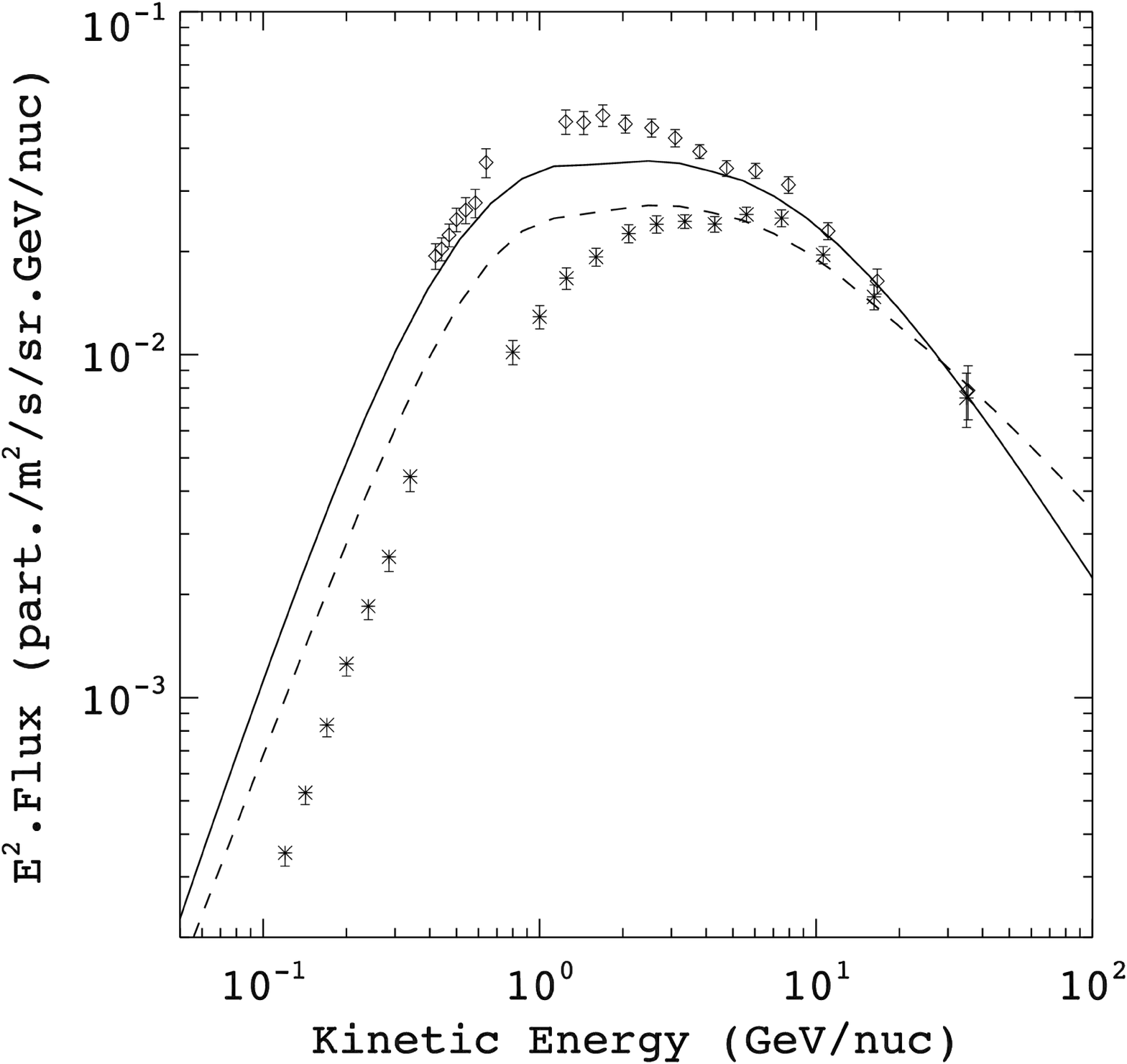}
             }
   \caption{Same as Fig. \ref{pri2} but the LIS for Secondary CR component species Manganese and Phosphorus is shown.}
   \label{sec3}
\end{figure}

\begin{figure}[!ht]
   \centerline{\includegraphics[width=2.8in, trim=1 15 1 27, clip=true]{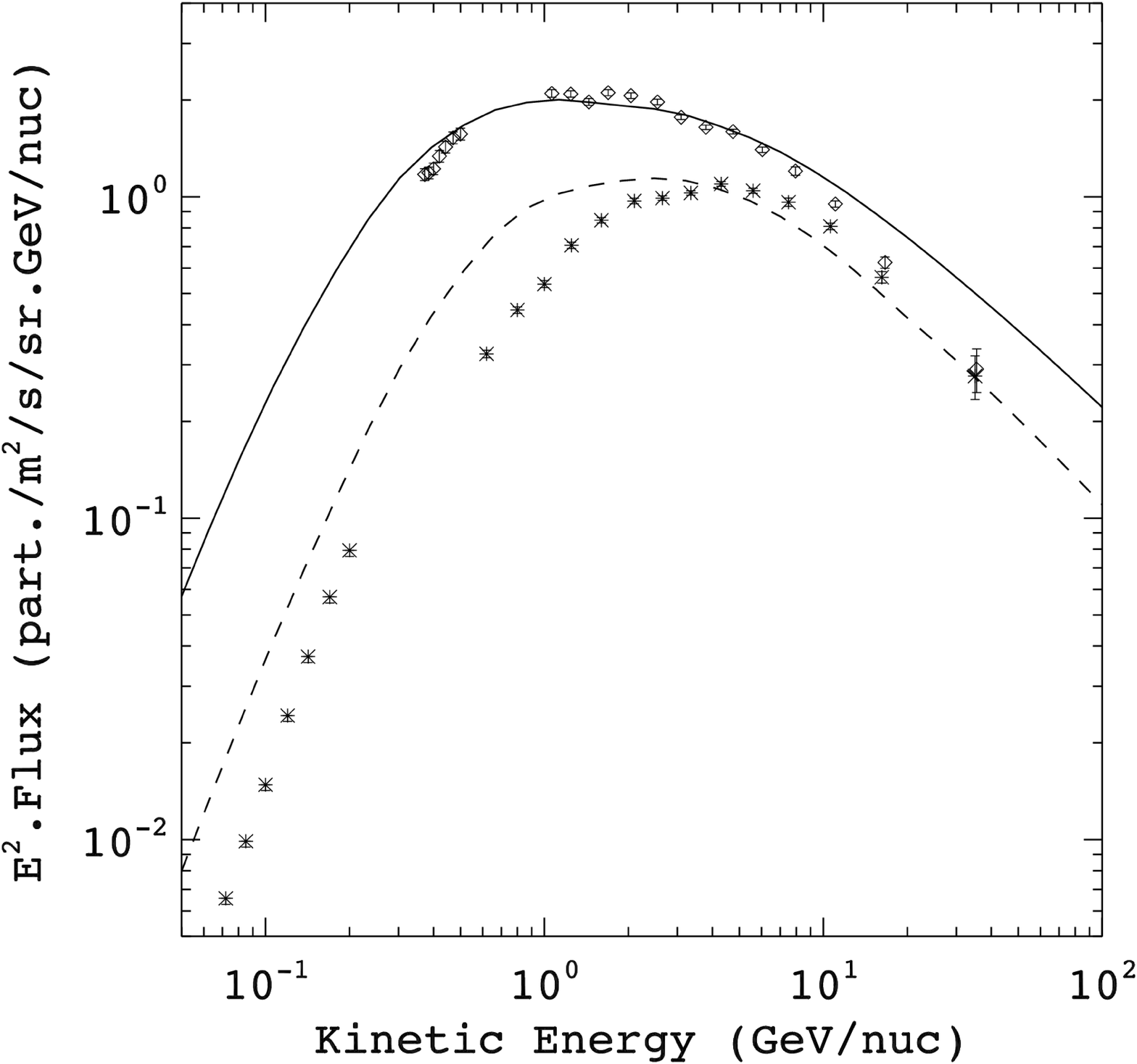}
              \hfil
              \includegraphics[width=2.8in, trim=1 15 1 27, clip=true]{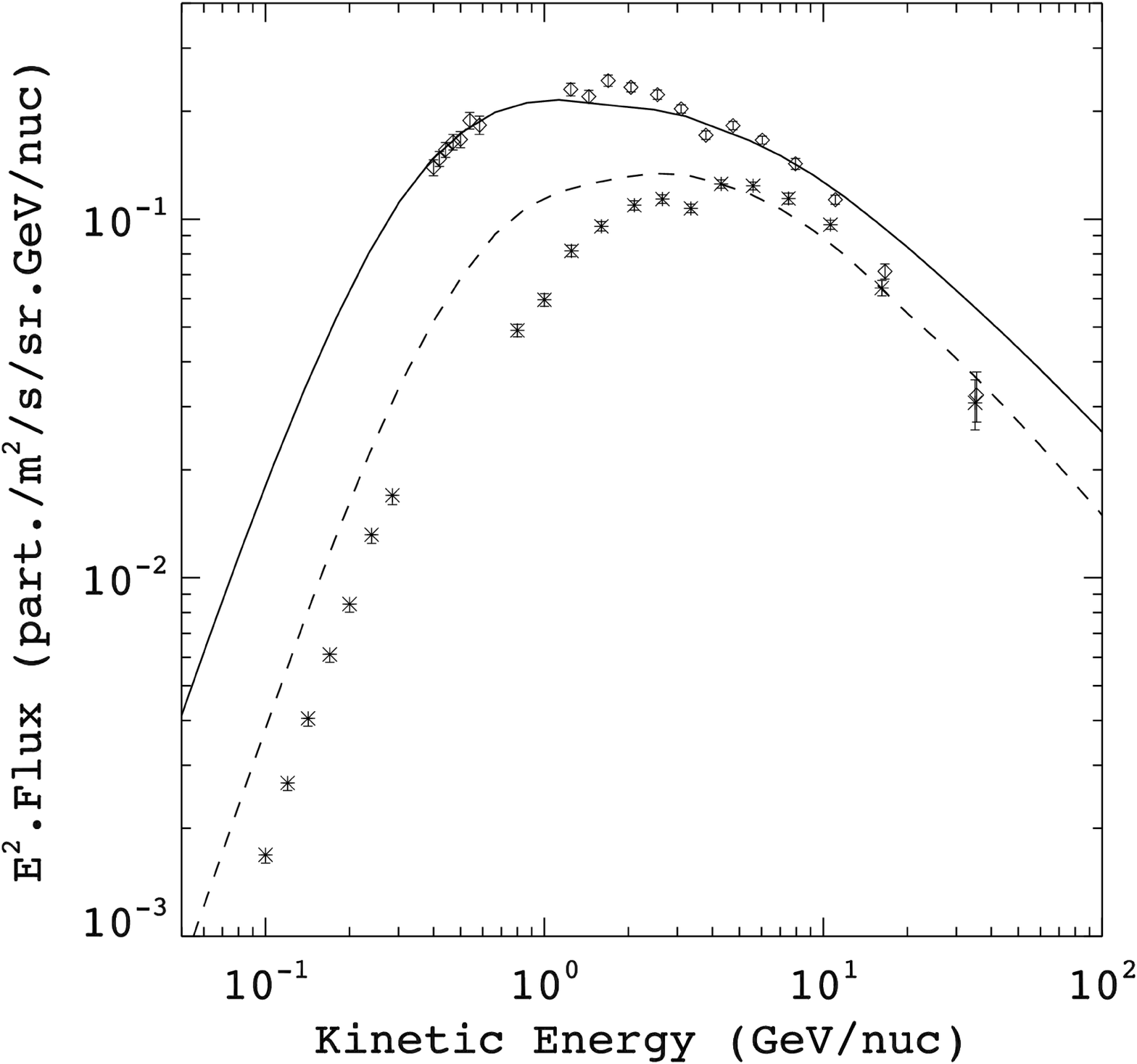}
            }
  \caption{Same as Fig. \ref{pri2} but the LIS for Mixed CR component species Nitrogen and Sodium is shown.}
   \label{mix2}
   \centerline{\includegraphics[width=2.8in, trim=1 15 1 27, clip=true]{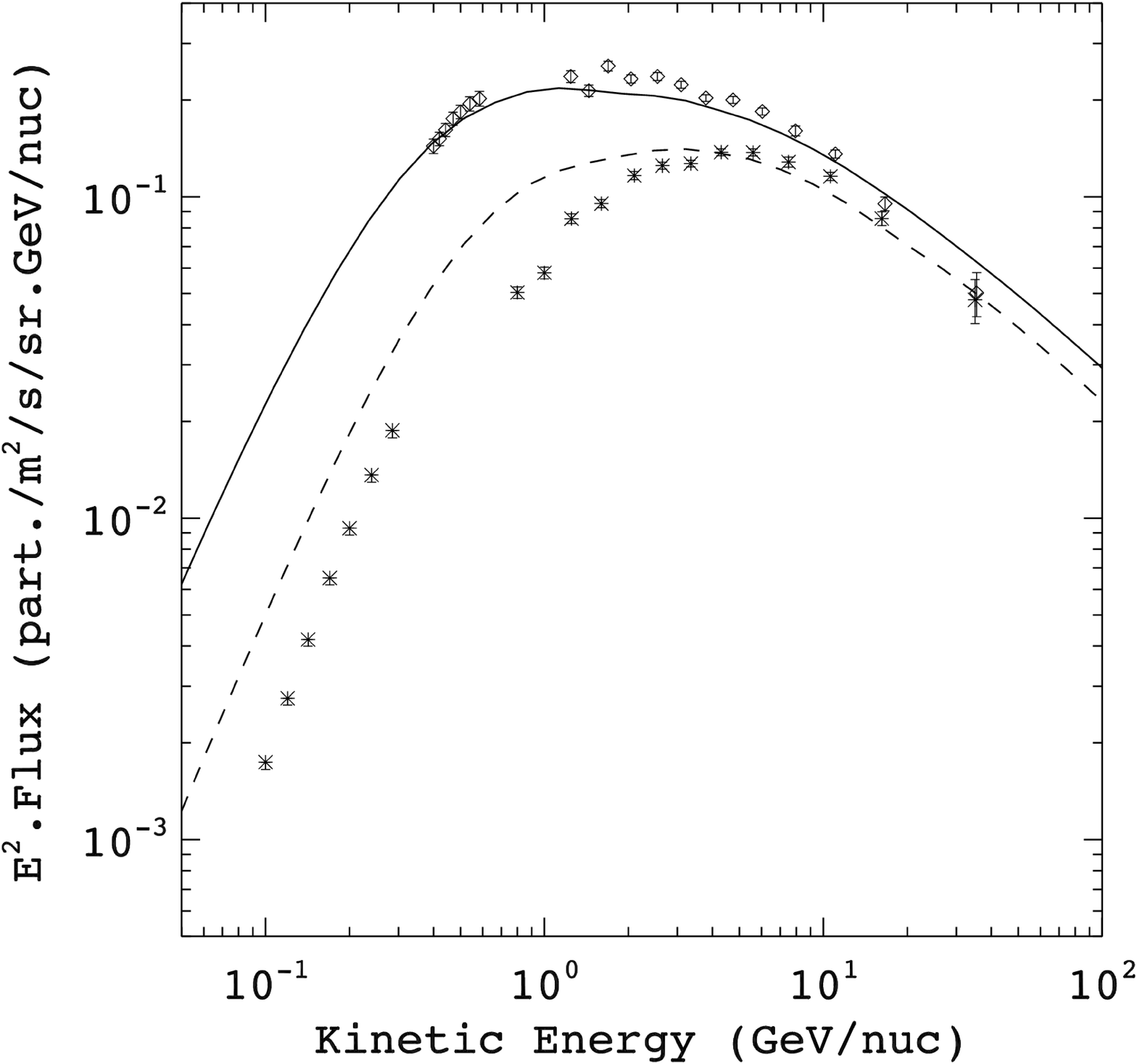}
              \hfil
              \includegraphics[width=2.8in, trim=1 15 1 27, clip=true]{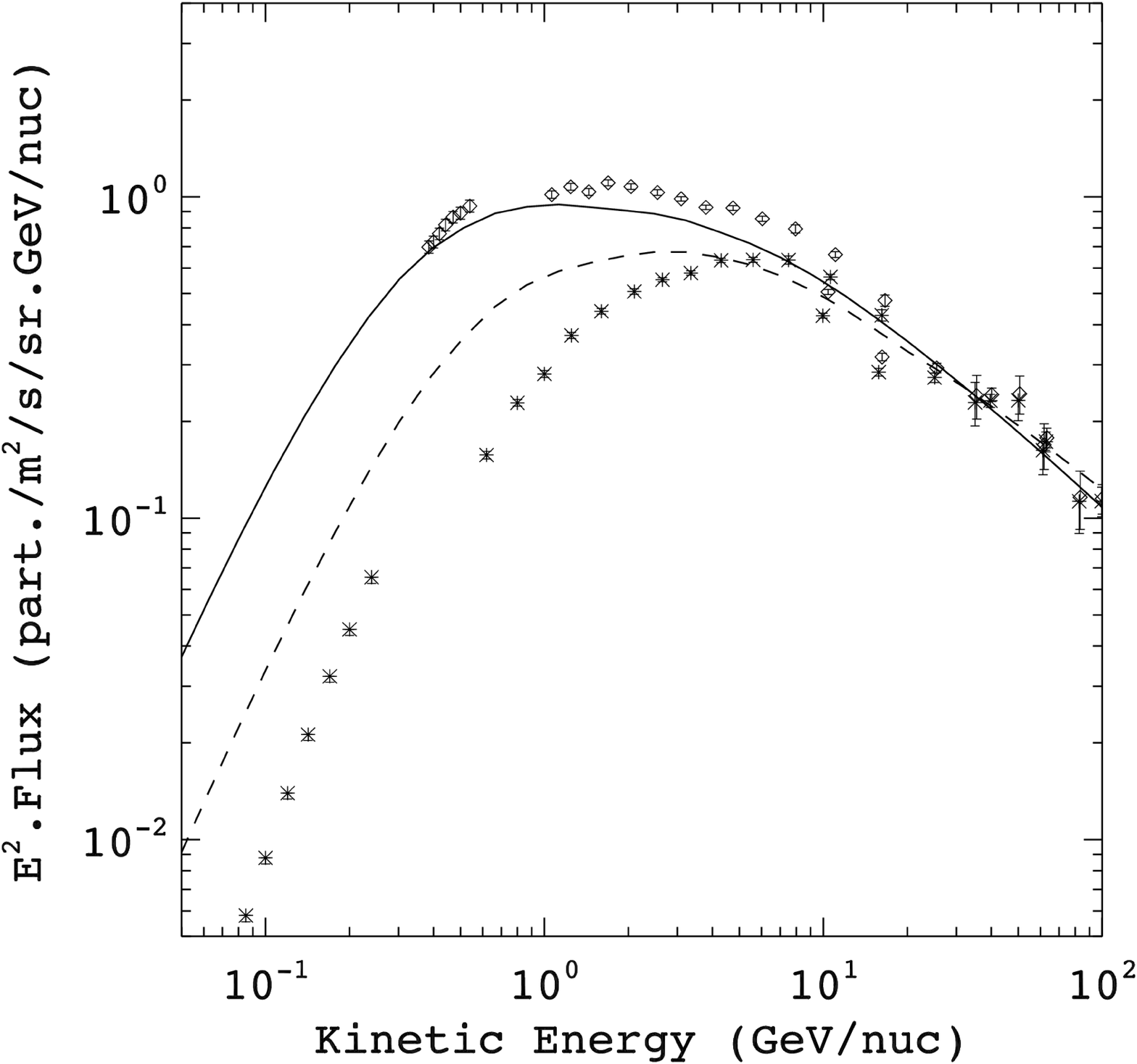}
            }
  \caption{Same as Fig. \ref{pri2} but the LIS for Mixed CR component species Aluminium and Neon is shown.}
   \label{mix3}
\end{figure}

The LIS produced by the best fit models with parameters listed in Table \ref{result}, are plotted in Fig. \ref{pri2} to \ref{mix3}. These figures show the LIS for selected CR species of the primary and secondary component groups. LIS obtained by \citet{Ptuskin2006} using the GALPROP code (but with a different analysis than the study conducted here) are also shown for comparison, because their parameter choices where used as a starting point for this parameter study. The LIS shown are Carbon and Iron for the primaries; Boron and Fluorine for the secondaries; and finally Nitrogen and Sodium for the mixed group. The experimental data and the corresponding demodulated data above 4\,GeV/nuc used to calculate the $\chi^2$ values are also shown.

All the  LIS presented by \citet{Ptuskin2006} are much lower than those LIS obtained in this study at energies below 10\,GeV/nuc. For the LIS shown, the \citet{Ptuskin2006} LIS do correspond to the obtained LIS at higher energies. The LIS for the primary CR species lie within the trend displayed by the data, with Iron lying in the lower part of the trend.  The LIS for the secondary CR species show good fits at all energies.

Small deviations in the fit of any one CR species in a group are to be expected due to the fact that all the calculated LIS of the CR species in a component group were simultaneously fitted to the data. The individual fitting of a CR species LIS may thus be lower or higher than expected to fit the data points, but for the whole group the $\chi^2$ value is still a minimum value. 

In this study, we aimed to include as many datasets as possible. We thus have to accept inconsistencies especially in the primary CR data due to systematic errors in the different experiments. This results in a wider spread of data points and thus larger $\chi^2$ values for species such as Iron, even though our best fit LIS can be seen to lie within the trend displayed by the data. Different experiments are not always consistently done and makes fitting the calculated LIS to the data difficult for such large data sets using the $\chi^2$ test. The mixed component shows similarly large $\chi^2$ values even though this component's experimental data doesn't show  the same amount of inconsistencies and suggests that the wide spread of data points might not be a determining factor.

A comparison of $\alpha$ and $\delta$ values to values obtained in studies conducted by \citet{Maurin2002}, \citet{Putze2009,Putze2010} and \citet{Trotta2011} is shown in Table \ref{compare1}. \citet{Maurin2002} studied the relation $\alpha + \delta = 2.8$ with a 2D diffusive propagation code similar to GALPROP and found the preferred values for $\alpha$ to be $>$ 2.0 and for $\delta$ to be $<$ 0.6 - 0.7. \citet{Putze2009} used a Monte Carlo technique with a Leaky-Box model and, by keeping the relation $\alpha + \delta = 2.65$ fixed, found that $\alpha$ should have a value of about 2.14 to 2.17 and $\delta$ a value of about 0.55 to 0.6. The follow up study by \citet{Putze2010} found a value of $\delta = 0.65$ with $\alpha + \delta = 2.65$ fixed for a plain diffusion model. From the contours can be seen that for the primaries the above stated $\delta$ and $\alpha$ values give very large $\chi^2$ values, indicating poor fits for primaries using these standard values. This is also true for the secondaries to a 
lesser extent. \citet{Trotta2011} conducted a full Bayesian parameter estimation for the GALPROP code and found constraints for $\delta$ between 0.26 to 0.35, and for $\alpha$ of between 2.29 to 2.47. Our values fall outside these constraints and this difference may be due to their inclusion of a break in the injection spectra $\alpha$ at 10\,GV. A note worthy difference between our study and those of the above mentioned authors, is the experimental data used. We use a larger range of isotopes to fit the data, as seen in Table \ref{crlist}, but did not include fits to generally used secondary to primary CR ratios due to the nature of the study. They included reacceleration, whereas we believe it to be of less importance for our study, which too might be the cause for the discrepancy in the values.

\begin{table}[!ht]
\footnotesize
\centering
\caption{Best fit $\alpha$ and $\delta$ values for all three components (columns two to four) compared to those obtained by \citet{Maurin2002}, \citet{Putze2009,Putze2010}, \citet{Trotta2011}.}
\label{compare1}
\begin{tabular}[t]{ l  l  l  l  l  l  l  l}
\hline
  & Primary & Mixed & Secondary & Maurin 2002   & Putze 2009  & Putze 2010 & Trotta 2011\\ \hline
$\alpha$  & 2.66    & 2.79  & 2.21      & $>$ 2.0       & 2.14 - 2.17 & ---        & 2.29 - 2.47\\
$\delta$  & 0.10    & 0.10  & 0.77      & $<$ 0.6 - 0.7 & 0.55 - 0.6  & 0.65       & 0.26 - 0.35\\
\end{tabular}
\end{table}

\section{Summary and Conclusions}
We performed a parameter study using the steady-state, 2D plain diffusion model of the public available GALPROP code. Looking at the CR primary and secondary components separately, we found that these components favour different best fit values for the magnitude of the diffusion coefficient $k_0$, the source spectral index $\alpha$ and the spectral index of the energy dependence of the diffusion coefficient $\delta$. This result is expected for the case where there is a significant contribution of nearby point sources to the local CR flux, as the steady-state 2D plain diffusion model is a good model for the CR secondary component but a poor model for the CR primary component due to its inadequate description of transient point sources.

The Primary and Secondary components seem to favour different regions in the scanned parameter space, and have different best fit values. The secondary CRs were found to be more easily fitted to data than the Primary component or the Mixed component group. This implies that the 2D GALPROP model as used is better suited for secondaries than for primaries. Thus, when only comparing CR data to LIS of secondary CRs the model can be used successfully. These results, together with the manner in which the 2D model handles CR sources, imply that there may be local sources of CRs that, so far, are not being taken into account. Such local sources would necessitate the use of more involved models, capable of taking point-like sources into account, than the one used in this study.

Comparing the best fit LIS found in the parameter study to the ones found by \cite{Ptuskin2006}, their LIS are seen to be much lower at larger energies than those obtained in this study, but certain LIS do agree favourably at higher energies. The differences between the LIS obtained in this study and the Ptuskin LIS, which are produced by the standard GALPROP parameter set, can possibly be attributed to dependence of the fitting on the data sets. Using different data sets or excluding data from certain experiments will have a significant effect on the best fit LIS found. Also, the method of including modulation is important, as choosing a modulation parameter for the force field model can be done arbitrarily.

Comparison of $\alpha$ and $\delta$ values found with those reported by \citet{Maurin2002} and \citet{Putze2009,Putze2010} showed that the Secondary component values are closer to those of the other studies than the Primary and Mixed component values. Our best fit values fall outside the parameter range as calculated by \citet{Trotta2011} which could be attributed to their larger set of free parameters, including reacceleration, and our use of higher order CRs. The fact that we found closed $\chi^2$ contours and distinct $\chi^2$ minima shows the high level of sophistication in the GALPROP model and the power of the ansatz used in this code to include as much information in its model as possible. 

Improvements to the assumptions made in this study could be made. For example a better modulation implementation such as a 2D drift model, with which more data at lower energies could be included with confidence. The influence on our results of parameters such as halo height, galactic wind and reacceleration could be investigated. Inclusion and changes to these parametrs might change the numerical results obtained, but can not concievebly change the conclusions on the seperate modelling of primary and secondary CRs, or possiblity of local sources.

\section*{Acknowledgement}
The authors wish to thank the South African National Research Foundation (NRF) and the SA High Performance Computing Centre (CHPC) for partial financial support.

\bibliographystyle{elsarticle-num-names}
\biboptions{numbers,sort&compress}
\bibliography{Discrepancy}

\newpage

\clearpage

\end{document}